\documentclass[preprint,aps,prd,amsmath,amssymb]{revtex4}
\pdfoutput=1
\usepackage{graphicx}
\usepackage{color}
\newcommand{\beq}{\begin{eqnarray}}
\newcommand{\eeq}{\end{eqnarray}}

\newcommand{\bmp}{\noindent\begin{minipage}{16cm}}
\newcommand{\emp}{\end{minipage}\vskip 7mm} % 7mm untightened

\usepackage{dcolumn}% Align table columns on decimal point
\usepackage{bm}% bold math
\usepackage{bbm}
\usepackage{subfigure}
\usepackage{pxfonts}

\usepackage{epsfig}

\def\lsim{\mathrel{\rlap{\lower4pt\hbox{\hskip1pt$\sim$}}
    \raise1pt\hbox{$<$}}}                % less than or approx. symbol
\def\gsim{\mathrel{\rlap{\lower4pt\hbox{\hskip1pt$\sim$}}
    \raise1pt\hbox{$>$}}}                % greater than or approx. symbol

\newcommand{\drawsquare}[2]{\hbox{%
\rule{#2pt}{#1pt}\hskip-#2pt%  left vertical
\rule{#1pt}{#2pt}\hskip-#1pt%  lower horizontal
\rule[#1pt]{#1pt}{#2pt}}\rule[#1pt]{#2pt}{#2pt}\hskip-#2pt%  upper horizontal
\rule{#2pt}{#1pt}}% right vertical

\newcommand{\Yfund}{\raisebox{-.5pt}{\drawsquare{6.5}{0.4}}}%  fund
\newcommand{\Ysymm}{\raisebox{-.5pt}{\drawsquare{6.5}{0.4}}\hskip-0.4pt%
        \raisebox{-.5pt}{\drawsquare{6.5}{0.4}}}%  symmetric second rank
\newcommand{\Ythrees}{\raisebox{-.5pt}{\drawsquare{6.5}{0.4}}\hskip-0.4pt%
          \raisebox{-.5pt}{\drawsquare{6.5}{0.4}}\hskip-0.4pt% 
          \raisebox{-.5pt}{\drawsquare{6.5}{0.4}}}%  symmetric third rank
\newcommand{\Yasymm}{\raisebox{-3.5pt}{\drawsquare{6.5}{0.4}}\hskip-6.9pt%
        \raisebox{3pt}{\drawsquare{6.5}{0.4}}}%  antisymmetric second rank
\newcommand{\Ythreea}{\raisebox{-3.5pt}{\drawsquare{6.5}{0.4}}\hskip-6.9pt%
        \raisebox{3pt}{\drawsquare{6.5}{0.4}}\hskip-6.9pt
        \raisebox{9.5pt}{\drawsquare{6.5}{0.4}}}

\newcommand{\Yadjoint}{\raisebox{-3.5pt}{\drawsquare{6.5}{0.4}}\hskip-6.9pt%
        \raisebox{3pt}{\drawsquare{6.5}{0.4}}\hskip-0.4pt
        \raisebox{3pt}{\drawsquare{6.5}{0.4}}}%  SU(3) adjoint
\begin{document}
%%%%%%%%%%%%%%%%%%%%%%%%%%%%%%%%%%%%%%%%%%%%%%%%%%%%%%%%%%%%%%%%%%%%%%%%%%%
\title{\Large  \color{red} QCD  Dual
 }
\author{Francesco {\sc Sannino}$^{\color{blue}{\varheartsuit}}$}
\email{sannino@cp3.sdu.dk} 
\affiliation{
$^{\color{blue}{\varheartsuit}}${ CP}$^{ \bf 3}${-Origins}, 
%IFK \& IMADA, University of Southern Denmark, 
Campusvej 55, DK-5230 Odense M, Denmark.\footnote{{ C}entre of Excellence for { P}article { P}hysics { P}henomenology devoted to the understanding of the {Origins} of Mass in the universe. This is the new affiliation from September 1$^{st}$ 2009.}}
%%%%%%%%%%%%%%%%%%%%%%%%%%%%%%%%%%%%%%%%%%%%%%%%%%%%%%%%%%%%%%%%%%%%%%%%%%%%%%%%%%%%%%%%
\begin{flushright}
{\it CP$^3$- Orgins: 2009-6}
\end{flushright}
\begin{abstract}
We uncover a novel solution of the 't Hooft anomaly matching conditions for QCD. Interestingly in the perturbative regime the new gauge theory, if interpreted as a possible QCD dual, predicts the critical number of flavors above which QCD in the nonperturbative regime, develops an infrared stable fixed point. Remarkably this value is identical to the maximum bound predicted in the nonpertubative regime via the all-orders conjectured beta function for nonsupersymmetric gauge theories. 
\end{abstract}

\maketitle

\section {Introduction}
Strong dynamics poses a formidable challenge. For decades physicists have been working on several aspects associated to the strongly coupled regime of gauge theories of fundamental interactions such as Quantum Chromodynamics (QCD). 

Perhaps one of the most fascinating possibilities is that QCD or alike have magnetic dual gauge theories. In fact, in the late nineties, in a series of exceptional papers Seiberg  \cite{Seiberg:1994bz,Seiberg:1994pq} provided strong support for the existence of a consistent picture of such a duality within a supersymmetric framework. Supersymmetry is, however, quite  special and the existence of such a duality does not automatically imply the existence of a QCD dual. One of the most relevant  Seiberg's results, for the present purpose,  has been the identification of the boundary of the conformal window for supersymmetric QCD as function of the number of flavors and colors.
The dual theories proposed by Seiberg pass a set of mathematical consistency relations known as 't Hooft anomaly conditions \cite{Hooft}.  Another important tool has been the knowledge of the all-orders supersymmetric beta function \cite{Novikov:1983uc,Shifman:1986zi,Jones:1983ip}

Recently we provided several analytic predictions for the conformal window of nonsupersymmetric gauge theories using different approaches \cite{Sannino:2004qp,Dietrich:2006cm,Ryttov:2007cx}. 
It is natural, at this point, to start exploring the possible existence of a QCD dual theory in the hope that it helps to provide a consistent picture of the QCD phase diagram as function of number of colors and flavors.

Arguably the existence of a possible dual of QCD able to reproduce its infrared dynamics must match the 't Hooft anomaly conditions \cite{Hooft}. We will exhibit several solutions of these conditions for QCD. An earlier exploration already appeared in the literature \cite{Terning:1997xy}. These conditions are, per se, not sufficiently constraining to select a unique QCD dual. However we have found a new solution with the property that in the perturbative regime predicts the critical number of flavors above which QCD, in the electric variables, enters the conformal regime as predicted using the all-orders conjectured beta function for nonsupersymmetric gauge theories \cite{Ryttov:2007cx}.

\section{QCD Global Anomalies and Conformal Window} 
The underlying gauge group is $SU(3)$ while the
quantum flavor group is
\begin{equation}
SU_L(N_f) \times SU_R(N_f) \times U_V(1) \ ,
\end{equation}
and the classical $U_A(1)$ symmetry is destroyed at the quantum
level by the Adler-Bell-Jackiw anomaly. We indicate with
$Q_{\alpha;c}^i$ the two component left spinor where $\alpha=1,2$
is the spin index, $c=1,...,3$ is the color index while
$i=1,...,N_f$ represents the flavor. $\widetilde{Q}^{\alpha ;c}_i$
is the two component conjugated right spinor. We summarize the
transformation properties in the following table.
\begin{table}[h]
\[ \begin{array}{|c| c | c c c|} \hline
{\rm Fields} &  \left[ SU(3) \right] & SU_L(N_f) &SU_R(N_f) & U_V(1) \\ \hline \hline
Q &\Yfund &{\Yfund }&1&~~1  \\
\widetilde{Q} & \overline{\Yfund}&1 &  \overline{\Yfund}& -1   \\
G_{\mu}&{\rm Adj}   &1&1  &~~1\\
 \hline \end{array} 
\]
\caption{Field content of an SU(3) gauge theory with quantum global symmetry $SU_L(N_f)\times SU_R(N_f) \times U_V(1)$. }
\end{table}

The  global anomalies are associated to the triangle diagrams featuring at the vertices three $SU(N_f)$ generators (either all right or all left), or two 
$SU(N_f)$ generators (all right or all left) and one $U_V(1)$ charge. We indicate these anomalies for short with:
\begin{equation}
SU_{L/R}(N_f)^3 \ ,  \qquad  SU_{L/R}(N_f)^2\,\, U_V(1) \ .
\end{equation}
For a vector like theory there are no further global anomalies. The
cubic anomaly factor, for fermions in fundamental representations,
is $1$ for $Q$ and $-1$ for $\tilde{Q}$ while the quadratic anomaly
factor is $1$ for both leading to
\begin{equation}
SU_{L/R}(N_f)^3 \propto \pm 3 \ , \quad SU_{L/R}(N_f)^2 U_V(1)
\propto \pm 3 \ .
\end{equation}
 Several analytic predictions for the lower end of the conformal window for nonsupersymmetric gauge theories with matter transforming according to several $SU$,  $SO$ and $Sp$ representation have been made \cite{Sannino:2004qp,Dietrich:2006cm,Ryttov:2007cx,Sannino:2009aw,Ryttov:2009yw}.  Here we add another method which uses {\it exact} anomaly matching conditions and demonstrate that one of the earlier analysis, i.e. the one based on the all-orders beta function conjecture, leads to a perfect match  with the  {\it exact} results.

Recently we have conjectured an all-orders beta function which allows for a bound of the conformal window \cite{Ryttov:2007cx} of gauge theories for any matter representation.  Other approaches yield compatible results. In this paper we provide one more piece of evidence in support of our conjecture.

Consider an $SU(N)$  gauge group with $N_f$ Dirac flavors belonging to the representation $r$ of the gauge group. 
The conjectured beta function \cite{Ryttov:2007cx} is given in terms of the anomalous dimension of the fermion mass $\gamma=-{d\ln m}/{d\ln \mu}$ where $m$ is the renormalized mass. At the zero of the all-orders beta function one has
\begin{eqnarray}
 \frac{2}{11}T(r)N_f(r) \left( 2+ \gamma \right) = C_2(G) \ ,
\end{eqnarray}
The generators $T_r^a,\, a=1\ldots N^2-1$ of the gauge group in the
representation $r$ are normalized according to
$\text{Tr}\left[T_r^aT_r^b \right] = T(r) \delta^{ab}$ while the
quadratic Casimir $C_2(r)$ is given by $T_r^aT_r^a = C_2(r)I$. The
trace normalization factor $T(r)$ and the quadratic Casimir are
connected via $C_2(r) d(r) = T(r) d(G)$ where $d(r)$ is the
dimension of the representation $r$. The adjoint
representation is denoted by $G$. 
Hence, specifying the value of the anomalous dimensions at the IRFP yields the last constraint needed to construct the conformal window. Requiring the absence of negative norm states  at the conformal point requires  $\gamma < 2$ resulting in the {\it maximum} possible extension of the conformal window bounded from below by:
\begin{equation}
N_f(r)^{\rm BF} \geq \frac{11}{8} \frac{C_2(G)}{T(r)}  \qquad { \gamma =2}\ .
\end{equation}
Specializing to three colors and fundamental representation we find: 
\begin{equation}
N_f(r)^{\rm BF} \geq \frac{33}{4} = 8.25 \ , \qquad {\rm for~QCD~with}  \quad  { \gamma =2}  .
\end{equation}
The actual size of the conformal window can, however, be smaller than the one determined above without affecting the validity of the beta function. It may happen, in fact, that chiral symmetry breaking is triggered for a value of the anomalous dimension less than two. If this occurs the conformal window shrinks. The ladder approximation approach\cite{Appelquist:1988yc,{Cohen:1988sq},Appelquist:1996dq,Miransky:1996pd},  for example, predicts that chiral symmetry breaking occurs when the anomalous dimension is larger than one. Remarkably the all-orders beta function encompass this possibility as well \cite{Ryttov:2007cx}. In fact, it is much more practical to quote the value predicted using the beta function by imposing $\gamma =1$:
\begin{eqnarray}\label{One}
N_f(r) \geq  \frac{11}{6}  \frac{C_2(G)}{T({r})} \ ,\qquad {\gamma =1}  \ .
\end{eqnarray}
{}For QCD we have:
\begin{equation}
N_f(r)^{\rm BF} \geq   11 \ ,\qquad {\rm for~QCD~with} \quad { \gamma =1} \ .
\end{equation}
The result is very close to the one obtained using directly the ladder approximation, i.e.  $N_f \approx 4 N$, as shown in \cite{Ryttov:2007cx,Sannino:2009aw}.  

 Lattice simulations of the conformal window for various matter representations  \cite{Catterall:2007yx,Catterall:2008qk,
Shamir:2008pb,DelDebbio:2008wb,DelDebbio:2008zf, Hietanen:2008vc,Hietanen:2008mr,Appelquist:2007hu,Deuzeman:2008sc,Fodor:2008hn,DelDebbio:2008tv,DeGrand:2008kx,Appelquist:2009ty,Hietanen:2009az,Deuzeman:2009mh,DeGrand:2009et,Hasenfratz:2009ea} are in agreement with the predictions of the conformal window via the all-orders beta function.

{}It would be desirable to have a novel way to determine the conformal window which makes use of exact matching conditions.  
%By comparing the various methods one can infer the anomalous dimension to pick as boundary of the window.

\section{Dual Set Up} 
If a magnetic dual of QCD does exist one expects it to be weakly coupled near the critical number of flavors below which one breaks  large distance conformality in the electric variables. Determining a possible unique dual theory for QCD is, however, not simple given the few mathematical constraints at our disposal, as already observed in \cite{Terning:1997xy}. The saturation of the global anomalies is an important tool but is not able to select out a unique solution. We shall see, however, that one of the solutions, when interpreted as the QCD dual, leads to a prediction of a critical number of flavors corresponding exactly to the one obtained via the conjectured all-orders beta function.

 We seek solutions of the anomaly matching conditions for a gauge theory $SU(X)$ with global symmetry group $SU_L(N_f)\times SU_R(N_f) \times U_V(1)$  featuring 
{\it magnetic} quarks ${q}$ and $\widetilde{q}$ together with $SU(X)$ gauge singlet states identifiable as baryons built out of the {\it electric} quarks $Q$. Since mesons do not affect directly global anomaly matching conditions we could add them to the spectrum of the dual theory.  We study the case in which $X$ is a linear combination of number of flavors and colors of the type $\alpha N_f + 3 \beta$ with $\alpha$ and $\beta$ integer numbers. 

We add to the {\it magnetic} quarks gauge singlet Weyl fermions which can be identified with the baryons of QCD but massless. The generic dual spectrum is summarized in table \ref{dualgeneric}.
\begin{table}[h]
\[ \begin{array}{|c| c|c c c|c|} \hline
{\rm Fields} &\left[ SU(X) \right] & SU_L(N_f) &SU_R(N_f) & U_V(1)& \# ~{\rm  of~copies} \\ \hline 
\hline 
 q &\Yfund &{\Yfund }&1&~~y &1 \\
\widetilde{q} & \overline{\Yfund}&1 &  \overline{\Yfund}& -y&1   \\
A &1&\Ythreea &1&~~~3& \ell_A \\
S &1&\Ythrees &1&~~~3& \ell_S \\
C &1&\Yadjoint &1&~~~3& \ell_C \\
B_A &1&\Yasymm &\Yfund &~~~3& \ell_{B_A} \\
B_S &1&\Ysymm &\Yfund &~~~3& \ell_{B_S} \\
{D}_A &1&{\Yfund} &{\Yasymm } &~~~3& \ell_{{D}_A} \\
{D}_S & 1&{\Yfund}  &{\Ysymm} &  ~~~3& \ell_{{D}_S} \\
\widetilde{A} &1&1&\overline{\Ythreea} &-3&\ell_{\widetilde{A}}\\
\widetilde{S} &1&1&\overline{\Ythrees} & -3& \ell_{\widetilde{S}} \\
\widetilde{C} &1&1&\overline{\Yadjoint} &-3& \ell_{\widetilde{C}} \\
 \hline \end{array} 
\]
\caption{Massless spectrum of {\it magnetic} quarks and baryons and their  transformation properties under the global symmetry group. The last column represents the multiplicity of each state and each state is a  Weyl fermion.}
\label{dualgeneric}
\end{table}
The wave functions for the gauge singlet fields $A$, $C$ and $S$ are obtained by projecting the flavor indices of the following operator
\begin{eqnarray}
\epsilon^{c_1 c_2 c_3}Q_{c_1}^{i_1} Q_{c_2}^{i_2} Q_{c_3}^{i_3}\ ,
\end{eqnarray}
over the three irreducible representations of $SU_L(N_f)$ as indicated in the table \ref{dualgeneric}. These states are all singlets under the $SU_R(N_f)$ flavor group. Similarly one can construct the only right-transforming baryons $\widetilde{A}$, $\widetilde{C}$ and $\widetilde{S}$ via $\widetilde{Q}$. The $B$ states are made by two $Q$ fields and one right field $\overline{\widetilde{Q}}$ while the $D$ fields are made by one $Q$ and two $\overline{\widetilde{Q}}$ fermions. $y$ is the, yet to be determined, baryon charge of the {\it magnetic} quarks while the baryon charge of composite states is fixed in units of the QCD quark one.The $\ell$s count the number of times the same baryonic matter representation appears as part of the spectrum of the theory. Invariance under parity and charge conjugation of the underlying theory requires $\ell_{J} = \ell_{\widetilde{J}}$~~ with $J=A,S,...,C$ and $\ell_B = - \ell_D$. 

Having defined the possible massless matter content of the gauge theory dual to QCD we compute the $SU_{L}(N_f)^3$ and $SU_{L}(N_f)^2\,\, U_V(1)$ global anomalies in terms of the new fields:   
 \begin{eqnarray}
SU_{L}(N_f)^3 &\propto &  X + \frac{(N_f-3)(N_f -6)}{2}\,\ell_A + \frac{(N_f+3)(N_f +6)}{2}\,\ell_S + (N_f^2 - 9)\,\ell_C\nonumber \\ &&   +\,(N_f-4)N_f\,\ell_{B_A} + \,(N_f+4)N_f\,\ell_{B_S}  + \frac{N_f(N_f-1)}{2}\,\ell_{D_A} \nonumber \\ 
&&+\frac{N_f(N_f+1)}{2}\,\ell_{D_S}  = 3 \ ,  \\ 
 & &\\
SU_{L}(N_f)^2\,\, U_V(1) &\propto &  y\,X +3 \frac{(N_f-3)(N_f -2)}{2}\,\ell_A + 3\frac{(N_f+3)(N_f +2)}{2}\,\ell_S + 3 (N_f^2 - 3)\,\ell_C \nonumber \\ 
 &&  +\, 3(N_f-2)N_f\,\ell_{B_A} + \,3(N_f+2)N_f\,\ell_{B_S}  + 3\frac{N_f(N_f-1)}{2}\,\ell_{D_A}\nonumber \\ &&+3\frac{N_f(N_f+1)}{2}\,\ell_{D_S} =3  \ .   \end{eqnarray}
  The right-hand side is the corresponding value of the anomaly for QCD. 
 
 \section{A Realistic QCD Dual}
We have found several solutions to the anomaly matching conditions presented above. Some were found previously in \cite{Terning:1997xy}. Here we start with a new solution in which the gauge group is $SU(2N_f - 5N)$  with the number of colors $N$  equal to $3$. It is, however, convenient to keep the dependence on $N$ explicit. 
 \begin{table}[bh]
\[ \begin{array}{|c| c|c c c|c|} \hline
{\rm Fields} &\left[ SU(2N_f - 5N) \right] & SU_L(N_f) &SU_R(N_f) & U_V(1)& \# ~{\rm  of~copies} \\ \hline 
\hline 
 q &\Yfund &{\Yfund }&1& \frac{N(2 N_f - 5)}{2 N_f - 5N} &~~~1 \\
\widetilde{q} & \overline{\Yfund}&1 &  \overline{\Yfund}& -\frac{N(2 N_f - 5)}{2 N_f - 5N}&~~~1   \\
A &1&\Ythreea &1&~~~3& ~~~2 \\
B_A &1&\Yasymm &\Yfund &~~~3& -2\\
{D}_A &1&{\Yfund} &{\Yasymm } &~~~3& ~~~2 \\
\widetilde{A} &1&1&\overline{\Ythreea} &-3&~~~2\\
 \hline \end{array} 
\]
\caption{Massless spectrum of {\it magnetic} quarks and baryons and their  transformation properties under the global symmetry group. The last column represents the multiplicity of each state and each state is a  Weyl fermion.}
\label{dual}
\end{table}
The solution above corresponds to the following value assumed by the indices and $y$ baryonic charge in table \ref{dualgeneric}.
  \begin{eqnarray}
 X=2N_f - 5N\ , \quad \ell_{A}=2\ ,  \quad \ell_{D_A} = -\ell_{B_A} =2 \ , \quad   \ell_{S}=\ell_{B_S} = \ell_{D_S} =\ell_C =0 \ , \quad y = N\,\frac{2 N_f - 5}{2 N_f - 15} \ ,\nonumber \\
 \end{eqnarray}
 with $N =3$.  $X$ must assume a value strictly larger than one otherwise it is an abelian gauge theory. This provides the first nontrivial bound on the number of flavors: 
 \begin{equation}
 N_f > \frac{5N + 1}{2}  \ , \end{equation} 
which for $N=3$ requires $N_f> 8 $. 
 \subsection{Conformal Window from the Dual Magnetic Theory}
Asymptotic freedom of the newly found theory is dictated by the coefficient of the one-loop beta function :
 \begin{equation}
 \beta_0 = \frac{11}{3} (2N_f - 5N)  - \frac{2}{3}N_f \ . 
 \end{equation}
 To this order in perturbation theory the gauge singlet states do not affect the {magnetic} quark sector and we can hence determine  the number of flavors obtained by requiring the dual theory to be asymptotic free. i.e.: 
\begin{equation}
N_f \geq \frac{11}{4}N \qquad\qquad\qquad {\rm Dual~Asymptotic~Freedom}\ . 
\end{equation}
Quite remarkably this value {\it coincides} with the one predicted by means of the all-orders conjectured beta function for the lowest bound of the conformal window, in the {\it electric} variables, when taking the anomalous dimension of the mass to be $\gamma =2 $. We recall that for any number of colors $N$ the all orders beta function requires the critical number of flavors to be larger than: 
\begin{equation}
N_f^{BF}|_{\gamma = 2} = \frac{11}{4} N \ . 
\end{equation}
{}For N=3 the two expressions yield $8.25$ \footnote{Actually given that $X$ must be at least $2$ we must have  $N_f \geq 8.5$ rather than $8.25$}. We consider this a nontrivial and  interesting result lending further support to the all-orders beta function conjecture and simultaneously suggesting that this theory might, indeed, be the QCD magnetic dual. 
%We note that we could not find a solution of the 't Hooft anomaly conditions leading to a possible dual theory with $X$ a linear combination of the number of flavors and colors leading to an asymptotically free condition for the dual theory  consistent with the condition $N_f \geq 11N/3 $. This is the value of the lower bound of the conformal window for QCD consistent with an anomalous dimension of the mass equal to unity. 
%
%Now, let's suppose that $\gamma=1$, and we require asymptotic freedom in dual theory.  This requires $N_f \geq 11 N /3$.  A bit of algebra shows that it   is 
% not possible to find a solution satisfying these conditions with $X$ linear in $N_f$. If one drops the anomaly matching conditions one can still find a gauge  theory whose one-loop beta function changes sign exactly at the point where the all-orders beta function, with $\gamma =1 $, would predict the end of the  conformal window.  
 The actual size of the conformal window matching this possible dual corresponds to setting $\gamma =2$. {}We note that although for $N_f = 9$ and $N=3$ the magnetic gauge group is $SU(3)$ the theory is not trivially QCD given that it features new massless fermions and their interactions with massless mesonic type fields.

To investigate the decoupling of each flavor at the time one needs to introduce bosonic degrees of freedom. These are not constrained by anomaly matching conditions. Interactions among the mesonic degrees of freedom and the fermions in the dual theory cannot be neglected in the regime when the dynamics is strong. The simplest mesonic operator $M_i^{j} $ transforming simultaneously according to the antifundamental representation of $SU_L(N_f)$ and the fundamental representation of  $SU_R(N_f)$  leads to the following type of interactions for the dual theory: 
\begin{eqnarray}
L_{\rm M} & = & Y_{q\widetilde{q}} \,\,q \, \, M \, \, \widetilde{q } +
% Y_{A\widetilde{A}}\,\,  A\, MMM \, \widetilde{A} +Y_{A\widetilde{S}} \,\, A\, MMM\, \widetilde{S} + Y_{A\widetilde{C}} \,\,A\, MMM \, \widetilde{C} +   Y_{S \widetilde{S}}\,\,  SMMM\widetilde{S}  + \nonumber \\
%&+&Y_{S\widetilde{C}} \,\, S\, MMM\, \widetilde{C} + Y_{C \widetilde{C}} \,\,C\, MMM\,  \widetilde{C} +  Y_{A{D_A}}\,\,  A\, MM\,  \overline{D}_A +Y_{A{D_S}} \,\, A\, MM\, \overline D_S  +\nonumber \\
%&+&  Y_{S{D_A}}\,\,  S\, MM\, \overline{D}_A +Y_{S{D_S}} \,\, S\, MM\, \overline D_S +   Y_{C{D_A}}\,\,  C \, MM\, \overline{D}_A +Y_{C {D_S}} \,\, C\, MM\, \overline D_S +  \nonumber \\
%&+&
Y_{A {B_A}} \,\,A\, M \overline{B}_A+ Y_{C {B_A}} \,\,C \, M \overline{B}_A +  Y_{C {B_S}} \,\,C \, M \overline{B}_S + Y_{S {B_S}} \,\,S \, M \overline{B}_S + \nonumber \\
&+&  Y_{B_A {D_A}}\,\,  B_A \, M\, \overline{D}_A +Y_{B_A {D_S}} \,\, B_A \, M\, \overline D_S + Y_{B_S {D_A}}\,\,  B_S \, M\, \overline{D}_A +Y_{B_S{D_S}} \, B_S \, M\, \overline D_S  + {\rm h.c.} 
\end{eqnarray}
The coefficients of the various operators are matrices taking into account the multiplicity with which each state occurs. The number of operators drastically reduces if we consider only  the ones linear in $M$.  The dual quarks and baryons interact  via mesonic exchanges. We have considered only the meson field for the bosonic spectrum because is the one with the most obvious interpretation in terms on the electric variables.
One can also envision adding new scalars charged under the dual gauge group \cite{Terning:1997xy} and in this case one can have contact interactions between the magnetic quarks and baryons. We expect these operators to play a role near the lower bound of the conformal window of the magnetic theory where QCD is expected to become free. It is straightforward to adapt the terms above to any anomaly matching solution. 

In Seiberg's analysis it was also possible to match some of the operators of the magnetic theory with the ones of the electric theory. The situation for QCD is, in principle, more involved although it is clear that certain magnetic operators match exactly the respective ones in the electric variables. These are the meson $M$ and the massless baryons, $A$, $\widetilde{A}$, ...., $S$ shown in Table \ref{dualgeneric}.  The baryonic type operators constructed via the magnetic dual quarks have baryonic charge which is a multiple of the ordinary baryons and, hence, we propose to identify them, in the electric variables, with bound states of QCD baryons. 

The generalization to a generic number of colors is currently under investigation \cite{SanninoWIP}. It is an interesting issue and to address it requires the knowledge of the spectrum of baryons for arbitrary number of colors.  It is reasonable to expect, however , a possible nontrivial generalization to any number of odd colors \footnote{For an even number of colors the baryons are bosons and a the analysis must modify.}. 

  \section{Earlier solutions}
  It is worth comparing the solution above with the ones found already in the literature \cite{Terning:1997xy}.  These are: 
 \begin{eqnarray}
 X=N_f - 6\ , \quad \ell_{A}=\ell_{D_A} = -\ell_{B_A} =1 \ , \quad   \ell_{S}=\ell_{B_S} = \ell_{D_S} =\ell_C = 0 \ , \quad y = 3\,\frac{N_f - 2}{N_f - 6} \ ,
 \end{eqnarray}
corresponding to $\alpha=1$ and $\beta=-2$, when taking the {\it magnetic} quark flavor symmetry assignment as in table \ref{dualgeneric}.  However, assigning the magnetic quarks $q$ to the complex representation of $SU_L(N_f)$ one has also the solution:
\begin{eqnarray}
 X=N_f  + 6\ , \quad \ell_{S}=\ell_{D_S} = -\ell_{B_S} =1 \ , \quad   \ell_{A}=\ell_{B_A} = \ell_{D_A} =\ell_C= 0 \ , \quad y = -3\,\frac{N_f + 2}{N_f +6} \ .
 \end{eqnarray}
 Assuming the gauge group to be $SU(N_f\pm 6)$ the one loop coefficient of the beta function is: 
 \begin{equation}
\beta_0 = \frac{11}{3}(N_f  \mp 6 ) - \frac{2}{3}N_f \ ,
\end{equation}
where the sign corresponds to the two possibilities for $X$, i.e. $N_f \mp 6$ and we have only included the magnetic quarks \footnote{ A complex scalar gauged under the dual gauge group was also added in the dual spectrum of \cite{Terning:1997xy}. The scalar  transformed according to the fundamental representation of the left and right $SU(N_f)$ groups. The resulting $SU(N_f -6)$ gauge theory is never asymptotically free and for this reason we have not included this scalar in our discussion. }. The critical number of flavors where asymptotic freedom is lost, in the case of the $N_f-6$ gauge group, corresponds to $7.33$. On the other hand we have a stronger constraint from the fact that the gauge group must be at least $SU(2)$ and hence $N_f \geq 8$ while no useful constraint can be obtained for the $N_f + 6$ gauge group. It was argued in \cite{Terning:1997xy}
that by taking $SO$ or $Sp$ as possible gauge group rather than $SU$ one might increase the critical number of flavors to around $10$. However choosing $SO$ and $Sp$ rather than $SU$ implies that the global symmetry group is enlarged to $SU(2N_f)$ and hence it is not clear how one can still match the anomaly conditions, unless one assumes a simultaneous dynamical enhancement of the QCD global symmetries at the fixed point. 
\section{Conclusions}
We uncovered a novel solution of the 't Hooft anomaly matching conditions for QCD. We have shown that in the perturbative regime the new gauge theory, if interpreted as a possible QCD dual, predicts the critical number of flavors above which QCD in the nonperturbative regime, develops an infrared stable fixed point. The value is identical to the maximum bound predicted in the nonpertubative regime via the all-orders conjectured beta function for nonsupersymmetric gauge theories. Recent suggestions to analyze the conformal window of nonsupersymmetric gauge theories based on different model assumptions \cite{Poppitz:2009uq} are in qualitative agreement with the precise results of the all-orders beta function conjecture. It is worth noting that the combination $2N_f - 5N$ appears in the computation of the mass gap for gauge fluctuations presented in \cite{Poppitz:2009uq,Poppitz:2008hr}. It would be interesting to explore a possible link between these different approaches in the future. 

Interestingly the present solution of the anomaly matching conditions indicate a substantial larger extension of the conformal window than the one predicted using the ladder approximation \cite{Appelquist:1988yc,Cohen:1988sq,Appelquist:1996dq,Miransky:1996pd} and the thermal count of the degree of freedom \cite{Appelquist:1999hr}. The prediction is, however, entirely consistent with the maximum extension of the conformal window obtained using the all-orders beta function \cite{Ryttov:2007cx}  together with a value of the anomalous dimension of the quark mass close to two. Our main conclusion is that the  't Hooft anomaly conditions alone do not exclude the possibility that the maximum extension of the QCD conformal window is the one obtained for a large anomalous dimension of the quark mass.  

By computing the same gauge singlet correlators in QCD and its suggested dual, one can directly validate or confute this proposal via lattice simulations.

\subsection*{Aknowledgments}
I gratefully thank T. Appelquist,  S. Catterall, L.~Del Debbio,  M.~T. Frandsen, J.~Giedt, C.~Pica, E.~Poppitz, T.~A. Ryttov, J.~Schechter, M.~Teper, L.G.~Yaffe,  M.~Unsal, A.~I.~Vainshtein and R.~Wijewardhana for comments on the manuscript or relevant discussions.

\end{document}